\pdfoutput=1
\documentclass[aps,prb,twocolumn,noshowpacs,a4paper,superscriptaddress]{revtex4-1}

\usepackage{graphicx}
\usepackage{color}
\usepackage{textcomp}
\usepackage{amssymb}\usepackage{amsmath}\usepackage{amsthm}
\usepackage{units}
\usepackage{braket}
\usepackage{bm}
\usepackage{bbold}

\begin{document}

\title{Equation of state of the two-dimensional Hubbard model}

\author{Eugenio Cocchi}
\thanks{These authors contributed equally to this work.}
\affiliation{Physikalisches Institut, University of Bonn, Wegelerstrasse 8, 53115 Bonn, Germany}
\affiliation{Cavendish Laboratory, University of Cambridge, JJ Thomson Avenue, Cambridge CB3 0HE, United Kingdom}
\author{Luke A. Miller}
\thanks{These authors contributed equally to this work.}
\affiliation{Physikalisches Institut, University of Bonn, Wegelerstrasse 8, 53115 Bonn, Germany}
\affiliation{Cavendish Laboratory, University of Cambridge, JJ Thomson Avenue, Cambridge CB3 0HE, United Kingdom}
\author{Jan H. Drewes}
\affiliation{Physikalisches Institut, University of Bonn, Wegelerstrasse 8, 53115 Bonn, Germany}
\author{Marco Koschorreck}
\affiliation{Physikalisches Institut, University of Bonn, Wegelerstrasse 8, 53115 Bonn, Germany}
\author{Daniel Pertot}
\affiliation{Physikalisches Institut, University of Bonn, Wegelerstrasse 8, 53115 Bonn, Germany}
\author{Ferdinand Brennecke}
\affiliation{Physikalisches Institut, University of Bonn, Wegelerstrasse 8, 53115 Bonn, Germany}
\author{Michael K{\"o}hl}
\email{michael.koehl@uni-bonn.de}
\affiliation{Physikalisches Institut, University of Bonn, Wegelerstrasse 8, 53115 Bonn, Germany}

\maketitle

\textbf{Understanding the phases of strongly correlated quantum matter is challenging because they arise from the subtle interplay between kinetic energy, interactions, and dimensionality. In this quest it has turned out that even conceptually simple models of strongly correlated fermions, which often only approximately represent the physics of the solid state, are very hard to solve\cite{Lee2006}. Since the conjecture by P.\,W. Anderson\cite{Anderson1987} that the two-dimensional Hubbard model describes the main features of high-T$_c$ superconductivity in the cuprates, there has been a major, yet inconclusive, research effort on determining its fundamental thermodynamic properties. Here we present an experimental determination of the equation of state of the repulsive two-dimensional Hubbard model over a broad range of interactions, $0\leq U/t \lesssim 20$, and temperatures, down to $k_BT/t=0.63(2)$, using high-resolution imaging of ultracold atoms in optical lattices. The equation of state fully characterizes the thermodynamics of the Hubbard model, and our results constitute benchmarks for state-of-the-art theoretical approaches.}

Ultracold fermionic atoms have emerged as a versatile platform to study strongly-correlated spin-1/2 fermions since they submit to a precise microscopic description and superbly sensitive detection. This approach has shed new light, for example, on the crossover between a Bose-Einstein condensate (BEC) of dimers and a Bardeen-Cooper-Schrieffer (BCS)-type superconductor as well as on the universal physics of the unitary Fermi gas\cite{Zwerger2012}. Among the remaining open questions are the properties of strongly-interacting fermions in lattices, which have begun to be explored\cite{Kohl2005b,Joerdens2008,Schneider2008,Taie2012,Hart2015,Duarte2015,Esslinger2010, Greif2015,Hofrichter2015}. However,  these investigations have not yet achieved the same level of accuracy in determining quantum phases and thermodynamic properties as those without lattice\cite{VanHoucke2012,Navon2010}. The experimental determination of the equation of state of the Hubbard model is of particular  importance because, even with the most advanced theoretical methods, strongly-correlated lattice models are notoriously hard to tackle\cite{Georges1996,Lee2006,Paiva2010}. Recently developed theoretical approximations of the two-dimensional Hubbard model\cite{Khatami2011,Leblanc2013}  provide predictions for a range of parameters, however, the inherent difficulty of simulating strongly-correlated fermions has yet precluded the determination of a general phase diagram, and the predictions resulting from the approximations still require experimental verification.

\begin{figure*}
\includegraphics[width=1.6\columnwidth,clip=true]{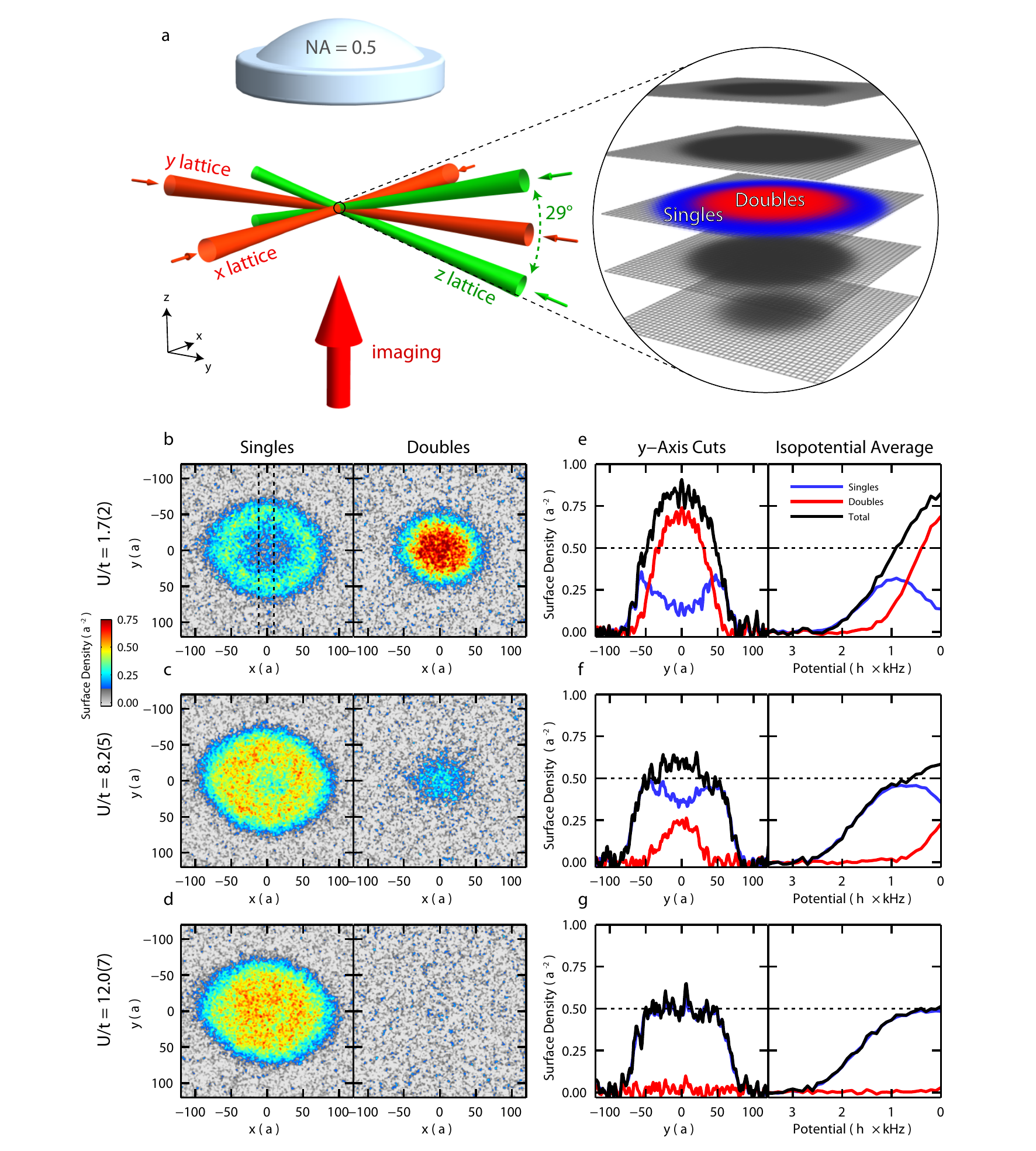}
\caption{Setup and in-situ density profiles. {\bf a} Experimental setup showing the arrangement of laser beams to create a stack of two-dimensional Hubbard models with ultracold atoms. Singles and doubles within a single layer are detected by combining radio-frequency spectroscopy and absorption imaging. {\bf b-d} In-situ density profiles of singles and doubles in the two-dimensional Hubbard model for different interaction strengths. The images are averaged over $\sim 35$ repetitions of the experiment. The density profiles cross over from a metallic phase at weak interactions, $U/t=1.6$, to a flat-top Mott insulator without doubles at strong interaction $U/t=12.0$. {\bf e-g} Singles, doubles, and total density data averaged over $-10a\leq x\leq 10a$ (see dashed lines in b) as well as evaluated along isopotential contours of the trapping potential.}
 \label{fig1}
\end{figure*}

The Hubbard model describes the two elementary processes of tunnelling between neighbouring lattice sites with amplitude $t$ and on-site interaction between two fermions of opposite spin  with strength $U$. In a single-band approximation the Hubbard Hamiltonian reads
\begin{align*}
H\!=\!-t \!\! \sum_{\braket{i,j},\sigma}\mkern-10mu\left(\hat{c}_{i,\sigma}^\dag \hat{c}_{j,\sigma}\!+\!\hat{c}_{j,\sigma}^\dag \hat{c}_{i,\sigma}\right)\!+U\!\sum_{i}\!\hat{n}_{i,\downarrow}\hat{n}_{i,\uparrow}-\mu\sum_{i,\sigma}\!\hat{n}_{i,\sigma}.
\end{align*}
Here $\hat{c}_{i,\sigma}$ ($\hat{c}_{i,\sigma}^\dag$) denotes the annihilation (creation) operator of a fermion on lattice site $i$ in spin state $\sigma=\{\uparrow, \downarrow\}$, the bracket $\braket{,}$ denotes the restricted sum over nearest neighbours,  $\hat{n}_{i,\sigma}=\hat{c}^\dag_{i,\sigma}\hat{c}_{i,\sigma}$ is the number operator and $\mu$ is the chemical potential. One of the key signatures of the repulsive ($U>0$) Hubbard model  is the appearance of a Mott insulating state at half filling, i.e. for $n\equiv (\braket{\hat{n}_{\uparrow}}+\braket{\hat{n}_{\downarrow}})/2=0.5$. The Mott insulator forms for $U\gg t, k_BT$, where $T$ is the temperature and $k_B$ is Boltzmann's constant. It is characterized by an occupation of one particle per lattice site and a gap against density excitations  of order $U$. 

Ultracold fermionic atoms in an optical lattice realize the Hubbard model\cite{Kohl2005b,Esslinger2010}. In such experiments, the Hamiltonian parameters  $t$, $U$, $\mu$, the temperature $T$, and the dimensionality are experimentally tunable, thus providing access to a large parameter range. Previously, investigations of the Hubbard model with ultracold atoms have mostly focused on the Mott insulator in three dimensions by detecting the global disappearance of doubly occupied sites\cite{Joerdens2008,Joerdens2010,Taie2012}, the response to an external compression\cite{Schneider2008}, the analysis of reconstructed density profiles\cite{Duarte2015}, and global detection of local spin correlations\cite{Greif2013,Hart2015}.  Unlike homogeneous solid state systems, ultracold atoms are  confined by an external trapping  potential $V(\bm{r})$ leading to a spatially varying density distribution $n(\bm{r})$. Therefore, different quantum phases can coexist in different regions of the trap and their unique identification using global observables is often impossible. Conversely, with sufficient local resolution, the coexistence of different phases can in principle be used to sample a range of the phase diagram in a single experimental realization. For bosonic\cite{Bakr2009,Gemelke2009,Sherson2010} and, recently, fermionic\cite{Duarte2015,Greif2015,Hofrichter2015} atoms in optical lattices the coexistence of different phases have been observed.

In this work we demonstrate high-resolution in-situ imaging of a spin-balanced  mixture of interacting spin-1/2 fermionic atoms in a single, two-dimensional layer of an optical lattice (see Figure 1a). By combining radio-frequency spectroscopy and absorption imaging we detect the in-situ density distributions of singly occupied lattice sites (``singles''),  $\braket{\hat{n}_\uparrow-\hat{n}_\uparrow \hat{n}_\downarrow}$, and of doubly occupied lattice sites (``doubles''), $\braket{\hat{n}_\uparrow \hat{n}_\downarrow}$ (for an extended description see Methods). This allows us to identify the two-dimensional Mott insulator and the metallic phase spatially resolved. Crucially, our technique gives direct access to the equation of state\cite{Ho2010} $n(\mu)$ and does not rely on  density reconstruction which can introduce numerical noise at small radii\cite{Duarte2015}. Our experiments cover the regimes from weak ($U/t \simeq 0$) to strong ($U/t \simeq 20$) interactions and, where available, we compare to state-of-the-art theories.

\begin{figure*}
 \includegraphics[width=1.9\columnwidth,clip=true]{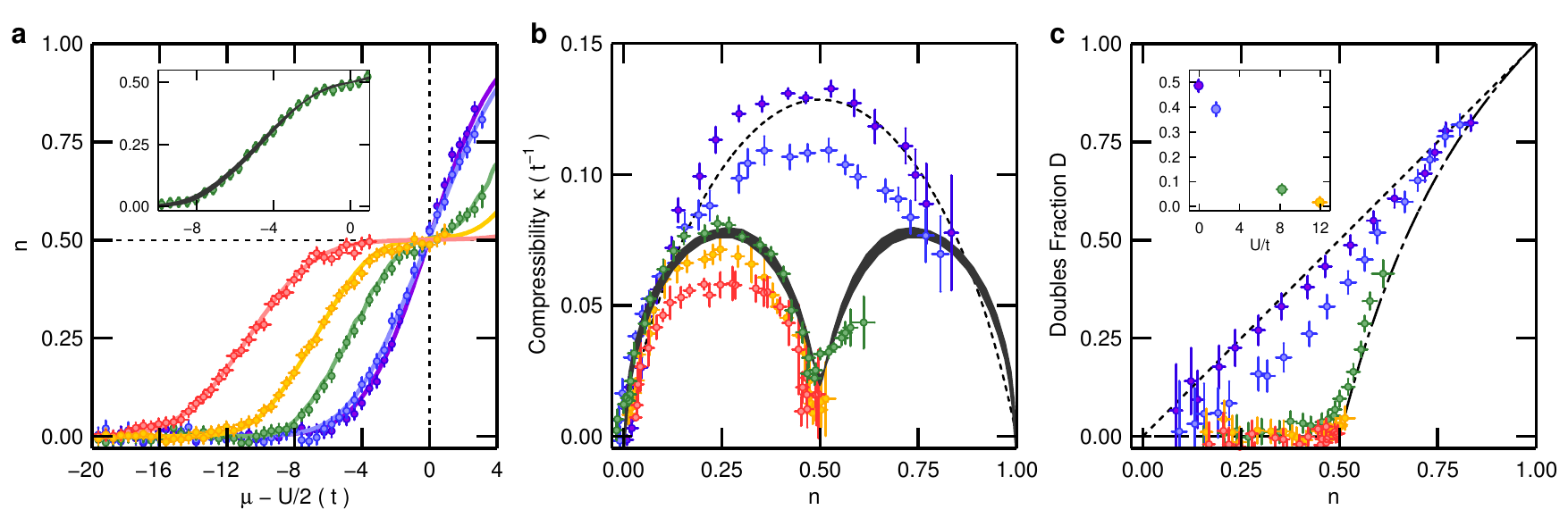}
 \caption{Equation of state of the two-dimensional repulsive Hubbard model. Purple: $U/t=-0.2(3)$ for $k_BT/t=1.35(4)$; blue: $U/t=1.6(2)$ for $k_BT/t=1.19(4)$; green: $U/t=8.2(5)$ for $k_BT/t=0.63(2)$; yellow $U/t=12.0(7)$ for $k_BT/t=0.92(6)$; red: $U/t=19.5(1.3)$ for $k_BT/t=1.41(5)$. {\bf a} Equation of state $n(\mu)$ vs. interaction strength.  Solid lines show fits using NLCE data\cite{Khatami2011} (purple: non-interacting Hubbard model) from which the temperature has been extracted. The inset shows the comparison with DCA data\cite{Leblanc2013} for the temperature interval $0.55\leq k_BT/t\leq 0.82$ at $U/t=8$. {\bf b} Compressibility $\kappa$ vs. filling. The dashed line shows the prediction of the non-interacting Hubbard model at $k_BT/t=1.4$ and the grey band the prediction from DCA as in (a). {\bf c} Doubles fraction vs. filling. The theoretical predictions for the non-interacting (dashed line) and infinitely repulsive (dashed-dotted line) Hubbard model are shown. The inset shows the behaviour at $n=0.5$ vs interaction strength. The error bars show the standard errors.}
 \label{fig2}
\end{figure*}

In Figure 1, we show examples of  in-situ density profiles for different interaction strengths (Fig. 1b-d) together with cuts through the density distribution (Fig. 1e-g). For weak interactions, $U/t=1.6(2)$, the system is metallic with an inhomogeneous density distribution. In the center of the trap the density is highest and we find an accumulation of doubles. This dense core is surrounded by a low-density  ring of singles (see Figure 1b). For intermediate interactions, $U/t=8.2(5)$, the  doubles are suppressed by the increased interaction energy and, correspondingly, the size of the cloud increases (see Figure 1c). Finally, for strong interactions, $U/t=12.0(7)$, we do not observe any doubles and a pronounced  plateau at filling $n=0.5$ forms, which signals the appearance of the Mott insulator (see Figure 1d). Employing the precise knowledge of the trapping potential $V(x,y)$ caused by the  envelope of the optical lattice beams (going beyond the harmonic approximation used in\cite{Gemelke2009,Ho2010,Duarte2015,VanHoucke2012,Navon2010}, see Methods) enables us to average the measured density along isopotential contours, see Figures 1e-g.

We analyse our data in the framework of the local density approximation, which states that the local chemical potential $\mu(x,y)$ results from the chemical potential at the center ($x=y=0$) of the cloud and the trapping potential  by $\mu(x,y)=\mu(0,0)-V(x,y)$, and that the properties of the homogeneous system can be locally applied.  The calibration of the central chemical potential $\mu(0,0)$ (based on the precise knowledge of $U$, see Methods) is provided by the particle-hole symmetry of the Hubbard model, according to which the maximum of singles occurs at  half filling,  $n=0.5$, where the chemical potential is $\mu(n=0.5)=U/2$. Combined with the knowledge of the potential this allows us to convert the recorded density profiles into an equation of state $n(\mu)$ and to directly determine  local thermodynamic properties of the gas\cite{Gemelke2009,Ho2010,Navon2010,VanHoucke2012}.

\begin{figure*}
 \includegraphics[width=1.2\columnwidth,clip=true]{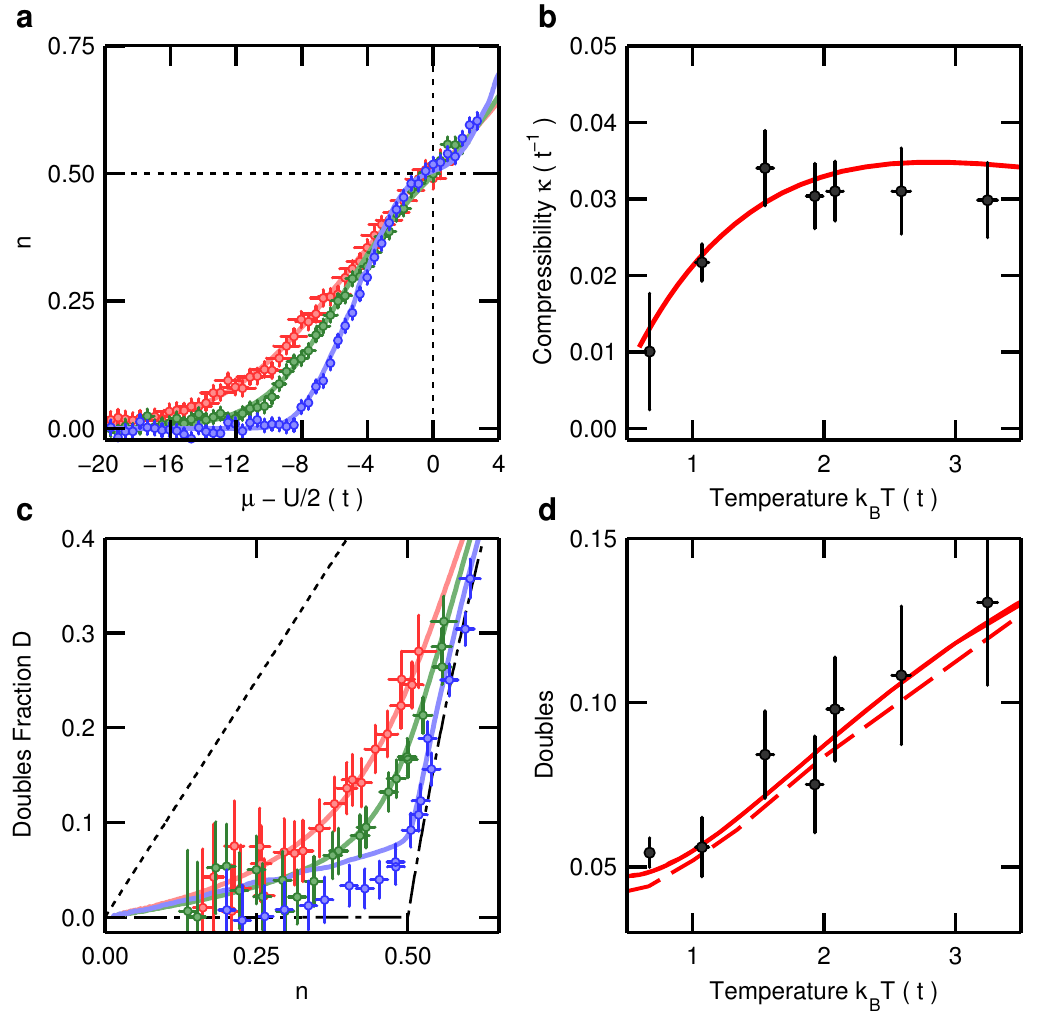}
 \caption{Equation of state vs. temperature at $U/t=8.2$. Blue: $k_BT/t=0.67(3)$, green: $k_BT/t= 1.55(6)$, red: $k_BT/t=3.25(7)$. {\bf a} Equation of state $n(\mu)$ for different temperatures with fits using NLCE data. {\bf b} Compressibility at half filling vs. temperature together with NLCE data.  {\bf c} Doubles fraction $D$ vs. filling for different temperatures. {\bf d} Doubles $\braket{\hat{n}_\uparrow \hat{n}_\downarrow}$ at half filling vs. temperature. The solid line shows the coinciding predictions of NLCE\cite{Khatami2011} and DCA\cite{Leblanc2013}, and the dashed line the QMC prediction\cite{Paiva2010}. The error bars show the standard errors.}
\end{figure*}

In Figure 2a, we show the experimental data of the equation of state  $n(\mu)$ for different interaction strengths. The data clearly show the crossover from a metallic phase for weak interactions with a strong density variation across the region of half filling, to a Mott-insulating phase with a plateau at half-filling for strong interactions. While the atom number remains approximately constant at $N\simeq 8.8(6)\times 10^3$ across different interaction strengths, there is a marked increase the extent of the Mott-insulating region for larger $U/t$. We also note that the filling $n=0.5$ at a chemical potential of $\mu=U/2$ is a fixed-point of the equation of state, as theoretically expected. We fit the measured equation of state with data from numerical linked cluster expansion (NLCE) calculations\cite{Khatami2011}, from which we determine the temperature of the gas. Our lowest temperature, $k_BT/t=0.63(2)$ at $U/t=8.2$, is already at the limit of validity of the numerical approximation, which is evident from the weak artificial oscillations of the theoretical $n(\mu)$ data near $n=0.25$ and above $n=0.5$. We also compare  our data  with  dynamical cluster approximation (DCA)\cite{Leblanc2013} for $U/t=8$ (see inset of Figure 2a), which confirms that they fall into the temperature interval $0.55\leq k_BT/t\leq 0.82$. However, the available DCA data are too coarsely spaced in order to fit the temperature more accurately.

From the equation of state  we compute the compressibility $ \kappa=\frac{\partial n}{\partial \mu}$ shown in Figure 2b. The non-interacting gas exhibits a maximum compressibility of $\kappa=0.133(3)\,t^{-1}$ at half filling $n=0.5$, which agrees with the numerical simulation  of the homogeneous non-interacting Hubbard model at temperature $k_BT/t=1.4$ (dashed line). The approach from the metallic phase to the Mott insulator is a crossover and therefore we expect a smooth change of the thermodynamic properties. For intermediate interactions, $U/t=8.2$, we already observe a significant reduction of the compressibility at $n=0.5$ and a return to a more compressible phase at fillings $n>0.5$. We compare this compressibility  with data from DCA calculations\cite{Leblanc2013} and find very good agreement. For strong interactions, $U/t\geq 12.0$, the compressibility is close to zero at half filling. For a system without disorder, a vanishing compressibility implies a gap against density (charge) excitations, and hence combined with the observation of a plateau at half-filling, unequivocally demonstrates the observation of the Mott insulator in two dimensions. 

Furthermore, in Figure 2c we show the measured doubles fraction $D=\frac{2\braket{\hat{n}_{\downarrow} \hat{n}_{\uparrow}}}{\braket{\hat{n}_\uparrow}+\braket{\hat{n}_\downarrow}}=\frac{\braket{\hat{n}_{\downarrow} \hat{n}_{\uparrow}}}{n}$ vs. filling and interaction strength.  For the non-interacting gas, the spin-up and spin-down fillings are uncorrelated,	 hence the doubles fraction simplifies to $D=n$ (dashed line in Fig. 2c).  The data for the non-interacting gas agree with this prediction. In the  limit of infinitely strong repulsive interactions, the number of doubles is completely suppressed if there are more lattice sites than particles, i.e. $D=0$ for $n\leq 0.5$, while for  $n>0.5$ the number of doubles  equals the excess of atoms above half filling, i.e. $D=2-1/n$ (dashed-dotted curve in Figure 2c).  We observe that even for  $U/t=8.2$ the data is close to the  infinite-interaction limit. For interaction strengths above this, the external compression provided by the trap is too weak to observe filling $n>0.5$. The inset shows the measured doubles fraction at half filling as a function of interaction strength $U/t$.

Finally, we investigate  the equation of state  for varying temperature for $U/t=8.2$. Around this interaction strength antiferromagnetic ordering is expected to occur at the highest transition temperature\cite{Paiva2010} and thus knowledge of the thermodynamics will guide the approach to this state. In order to experimentally adjust the temperature, we heat the gas using a weak periodic modulation of the intensity of the horizontal lattice  beams with a frequency close to twice the horizontal trapping frequency, followed by an equilibration time (see Methods). In Figure 3a, we display how $n(\mu)$ varies with temperature. The distribution $n(\mu)$ gets broader and the compressibility at half-filling increases. The latter is detailed in Figure 3b which shows how the Mott insulator melts and its compressibility increases from $\kappa=0.01\,t^{-1}$ at low temperature to $\kappa=0.03\,t^{-1}$ at high temperature, in agreement with NLCE data (solid line). Moreover, in Figure 3c we show the doubles fraction $D$  vs. filling $n$ for varying temperatures and find a considerable increase of doubles with temperature across all fillings. Some deviations between experimental and theoretical NLCE data are observed at low filling for low temperatures even though the total filling $n$ is in good agreement. In Figure 3d we plot  the doubles  at half filling vs. temperature and find that it increases from $0.054(5)$ at the lowest temperatures $k_BT/t =0.67(3)$ to $ 0.13(3)$ at $k_BT/t=3.25(7)$, in agreement with the results from  NLCE \cite{Khatami2011}, DCA\cite{Leblanc2013}, and quantum Monte-Carlo (QMC) simulations\cite{Paiva2010}. Both, the increase of compressibility and the increase in doubles,  signal the creation of thermally activated density excitations out of the lower Hubbard band. For the lowest temperatures $k_BT/t=0.63$, our data are, within error, in agreement with the zero-temperature extrapolation of numerical theory calculations. This reflects that our measurements have reached the temperature limit for which density-dependent quantities are suited for studies of the thermodynamics of the Hubbard model, and we consider our measurements complete for the density (charge) sector.

\section{Methods summary}
We start from a two-component, spin-balanced Fermi gas of $^{40}$K atoms in the two lowest hyperfine ground states, $\ket{\downarrow}\equiv \ket{F=9/2,m_F=-9/2}$ and $\ket{\uparrow}\equiv \ket{F=9/2,m_F=-7/2}$, which is evaporatively cooled in a crossed-beam optical dipole trap to a temperature of $T=0.12(2)\,T_F$ with $1.2(1)\times 10^5$ atoms per spin state. Here, $F$ and $m_F$ denote the hyperfine quantum numbers, and $T_F$ is the Fermi temperature in the harmonic trap. We split the cloud into a stack of $\sim 11$ separate two-dimensional planes by ramping up a strong optical lattice potential along the vertical ($z$--) direction with a period of $a_z=1.06(1)\,\mu$m and a depth of $119(1) E_\text{rec,z}$, where $E_\text{rec,z} = h^2/(8 m a_z^2)$, $h$ is Planck's constant, and $m$ is the atomic mass. The  tunnelling between adjacent planes and the occupation of higher bands is negligible. The two-dimensional gases are then subject to an in-plane square optical lattice, realizing the two-dimensional Hubbard model. The lattice has a period of $a= 532$\,nm and a depth of $6.0(1)\,E_{\text{rec}}$ where $E_{\text{rec}}=h^2/(8ma^2)$, resulting in a tunnelling matrix element of $t=h\times 224(6)$\,Hz. The global confinement is, to lowest order, harmonic with trapping frequencies $\omega_x=2 \pi \times 22.0(2)$\,Hz and $\omega_y=2\pi\times 26.6(3)$\,Hz. We independently control the interaction parameter $U$ using the Feshbach resonance between the $\ket{\uparrow}$ and $\ket{\downarrow}$ states at a magnetic field of 202\,G. We use magnetic fields between 189\,G and 212\,G to tune the interactions over the range $0\lesssim U/t\leq 20$. To observe the atomic density  in a single horizontal plane we ramp up the horizontal lattices to $\simeq 60\,E_{\text{rec}}$ in order to freeze the density distribution and perform high-resolution radio-frequency (RF) spectroscopy in a pulsed, vertical magnetic field gradient. This is followed by RF spectroscopy in a homogeneous magnetic field to discriminate between singles and doubles  via the on-site interaction shift. Finally, we take consecutive absorption images of the two distributions.

\section{Methods}

\subsection*{Detection}

During the detection sequence we suppress atomic tunnelling by increasing the depths of the horizontal lattice to $60E_\mathrm{rec}$ resulting in a tunnelling matrix element of $h\times \unit[0.038]{Hz}$. To transfer one of the spin components within a selected layer into a third hyperfine state we use high-resolution radio-frequency (RF) spectroscopy in a vertically oriented magnetic field gradient. Interaction shifts between singly and doubly occupied lattice sites are avoided by tuning the homogeneous magnetic field to a value of $\unit[213.7]{G}$ where the scattering lengths in the collisional channels $|F=9/2,m_F=-7/2\rangle \& |F=9/2,m_F= -3/2\rangle$ and $|F=9/2,m_F = -7/2\rangle \& |F=9/2,m_F = -5/2\rangle$ have been measured to be identical. The direction of the magnetic field gradient of magnitude $B' = \unit[33.3(5)]{G/cm}$ was carefully aligned (to within an accuracy of $\unit[0.3]{mrad}$) along the direction of the vertical lattice using magnetic compensation fields along the $x$- and $y$-directions. The amplitude shape of the RF pulse used for addressing a single layer was optimized in order to minimize transfer of neighbouring layers. As shown in the Extended Data Fig.~1 we clearly resolve individual layers with a frequency spacing of $\unit[0.64(1)]{kHz}$. From a comparison between the observed and the expected contrast of the tomography spectrum we estimate a standard deviation of residual frequency fluctuations of $\unit[0.14(1)]{kHz}$ which corresponds to a  standard deviation of magnetic field fluctuations of $\unit[0.8(1)]{mG}$. From this we deduce an average population admixture from both neighbouring planes of $\unit[2.6(3)]{\%}$ within a frequency window of $\pm \unit[100]{Hz}$ around the resonance frequency of the selected layer.
 
To distinguish singles and doubles within the selected layer, we subsequently perform RF spectroscopy at a homogeneous magnetic field of $\unit[180]{G}$ where, due to the presence of the second spin component in state $|F=9/2, m_F=-5/2\rangle$, the transition frequency from $|F=9/2,m_F = -9/2\rangle$ to $|F=9/2,m_F = -7/2\rangle$ is shifted for doubles by $\unit[-6.6]{kHz}$ compared to the bare transition frequency. We address this transition using a narrow RF sweep (HS1-pulse \cite{Garwood2001}) over a range of $\unit[4]{kHz}$. For successive imaging of singles and doubles, we temporarily transfer the singles population into the $|F=7/2,m_F=-7/2\rangle$ hyperfine state using adiabatic microwave frequency (MW) sweeps. 

The overall detection fidelity of singles and doubles was calibrated to be $1.00(2)$ and $0.82(2)$, respectively, which was accounted for in the data analysis. The reduced detection fidelity of doubles is attributed to residual loss in the vicinity of the $p$-wave Feshbach resonance between states $|F=9/2,m_F=-9/2\rangle$ and $|F=9/2,m_F=-5/2\rangle$ at a magnetic field of $\unit[215]{G}$. We experimentally confirmed that the sum of the individually recorded density distributions of singles and doubles agrees to within an uncertainty below $1\%$ with the total density distribution recorded in the absence of the singles/doubles spectroscopy.

\renewcommand{\figurename}{EXTENDED DATA FIG.}
\setcounter{figure}{0}
\begin{figure}
 \includegraphics[width=\columnwidth,clip=true]{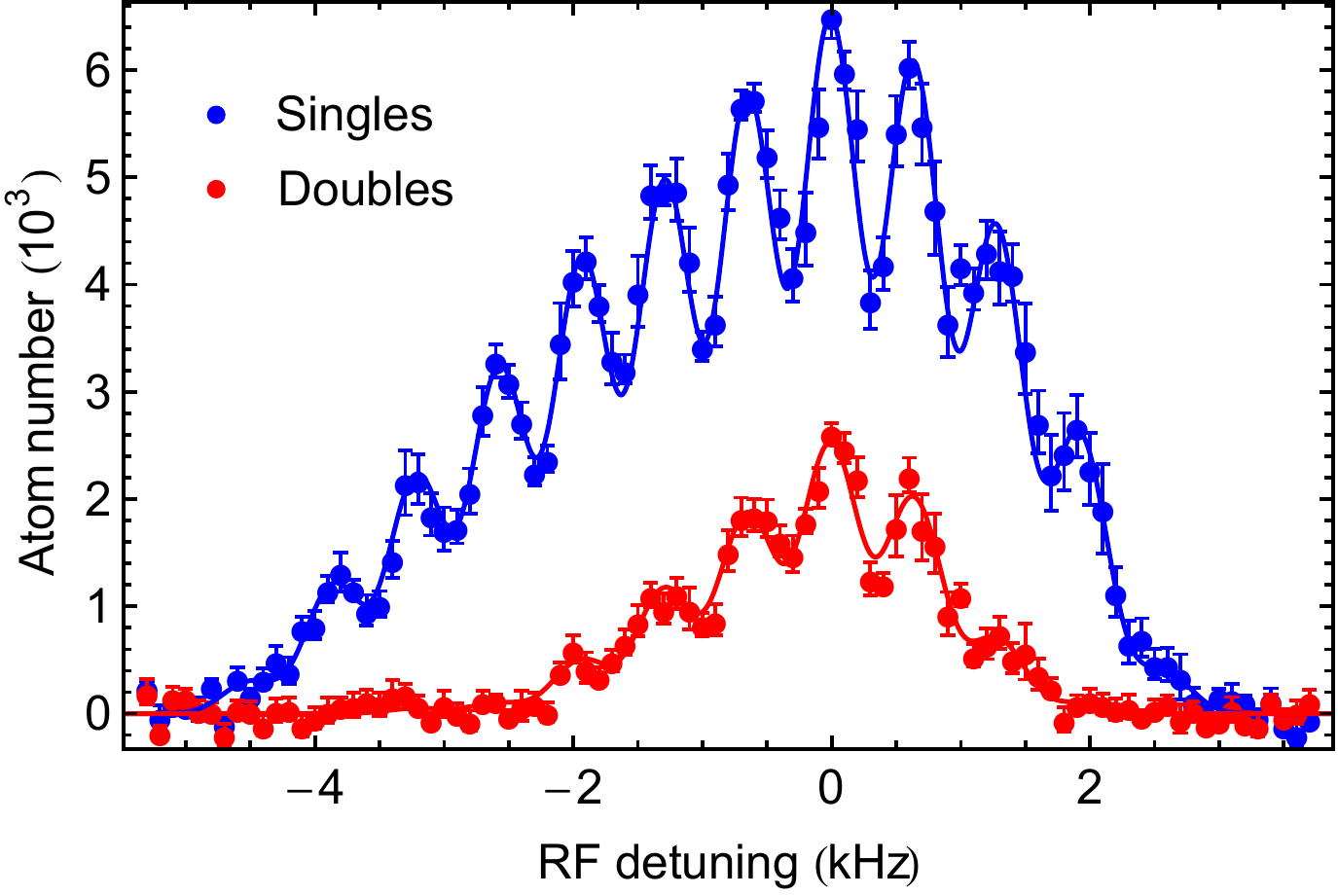}
 \caption{Vertical tomography. Shown is the total atom number on singly (blue) and doubly (red) occupied lattice sites obtained from RF tomography in a vertical magnetic field gradient of $B' = \unit[33.3]{G/cm}$. Data points show the average over nine experimental repetitions, the error bar displays the standard error. The solid lines are fits to the data based on the expected transfer function of the tomography RF-pulse taking into account a Gaussian residual frequency noise with a variance of $\unit[140]{Hz}$ and the zero-point width of the individual layers along the vertical direction. The agreement of the resonance frequencies for singles and doubles confirms the absence of an interaction shift at the chosen magnetic field bias of $\unit[213.7]{G}$.}
 \label{extendeddatafig2}
\end{figure}

\subsection*{Absorption imaging}
We perform absorption imaging on the $|F=9/2, m_F=-9/2\rangle$ to $|F'=11/2,m_F'= -11/2)$ cycling transition of the D2-line with a resonant optical pulse of $\unit[5]{\mu s}$ duration. The shadow cast by the atoms is imaged onto a CCD camera using an objective with numerical aperture of NA=0.5 and overall magnification of $22.7(1)$ which was calibrated from noise correlation measurements after ballistic expansion \cite{Rom2006}. The imaging resolution (FWHM of the autocorrelation peak) of $\unit[1.2(1)]{\mu m}$ was deduced from the density noise of individual in-situ images.

To account for saturation effects during imaging, we carefully calibrated the effective saturation intensity $I_\mathrm{sat}^* = \alpha^* I_\mathrm{sat}$ at the position of the atoms, which deviates by a factor $\alpha^*$ from the bare saturation intensity $I_\mathrm{sat}$ due to polarization imperfections and frequency fluctuations of the imaging light, as well as recoil-induced Doppler shifts \cite{Reinaudi2007}.
A series of absorption images of a single layer (with maximum optical density of $\sim 0.4$) was taken for varying imaging pulse durations $\tau = \unit[1.5 - 10]{\mu s}$ and approximately constant total photon number. The corresponding imaging light intensity $I_i$ varied between 1.2 and 7.1 times the saturation intensity $I_\mathrm{sat}$. This parameter range was chosen in order to limit the induced Doppler shift to below 0.7 of the intensity-broadened line width and the atomic displacement to about the depth of focus of the imaging system.

The optical depth evaluated from the set of calibration images
\begin{equation}
\mathrm{OD}_\mathrm{corr}(x,y) = -\alpha^* \log \left(\frac{I_f(x,y)}{I_i(x,y)}\right)+\frac{I_i(x,y)-I_f(x,y)}{I_\mathrm{sat}}
\end{equation}
where $I_i$ ($I_f$) denotes the initial (final) light intensity, respectively, was found to show least variation with imaging light intensity for a correction factor of $\alpha^* = 1.6(2)$. The saturation intensity $I_\mathrm{sat}$ was calibrated with an estimated accuracy of $10\%$ using a power meter. From the corrected optical density profile, the atomic density distribution $n(x,y)$ is deduced according to $n(x,y) = \mathrm{OD}_\mathrm{corr}(x,y)/\sigma_0$ with the bare cross section $\sigma_0 = 3 \lambda^2/2 \pi$ and the imaging wavelength $\lambda = \unit[766.7]{nm}$. The estimated systematic uncertainty of the thereby calibrated atomic density profiles is about 10\%, however, they provide the expected density of half-filling for a Mott insulator without applying an additional correction factor.

We confirmed by varying the duration of the imaging pulse that loss of doubles caused by light-assisted collisions can be neglected for our imaging parameters. The detection fidelity of doubles in absorption imaging was observed to be equal both, for in-situ imaging and imaging after ballistic expansion.

\subsection*{Imaging background}
Off-resonant imaging of the spin component residing in the hyperfine state $|F=9/2, m_F = -5/2\rangle$ causes a small background signal in the absorption images of the singles and doubles densities. We correct for this by subtracting from the individual images an average of several background images taken with a far off-resonant tomography pulse. From a comparison between the magnitude of the measured and expected imaging background, we deduce a lower bound of the average efficiency of the individual adiabatic RF sweeps of $0.998$.

\subsection*{Calibration of trapping potential}
In accordance with the experimental setup, we model the trapping potential $V(x,y)$ along the horizontal directions with a pair of Gaussian beams intersecting in the $y$-$z$-plane at an angle of $14.5(1)^\circ$ ($z$-lattice) and a combination of two in-plane, retro-reflected Gaussian beams ($x$/$y$-lattice). The power ratios between the incoming and retro-reflected beams are set to the measured values. We deduced the orientation of the horizontal lattice beams in the imaging frame from density-density correlations of a non-interacting gas imaged after ballistic expansion \cite{Rom2006}. The waists of the Gaussian beams are extracted from trap frequency measurements  performed in the absence of the retro-reflected beam along the oscillation direction. In order for the measurement to be dominated by the harmonic region of the Gaussian trapping potential, the oscillation amplitude and atom number in these measurements were sufficiently small. The measured trap frequencies are $\omega_x = 2 \pi \times \unit[22.0(2)]{Hz}$ and $\omega_x = 2 \pi \times \unit[26.6(3)]{Hz}$ which correspond to waists of $w_\mathrm{x-latt} = \unit[172(2)]{\mu m}$, $w_\mathrm{y-latt} = \unit[165(2)]{\mu m}$ and $w_\mathrm{z-latt} = \unit[140(5)]{\mu m}$ as was confirmed within errors by direct imaging of the beam profiles. 

\subsection*{Calibration of lattice depths and Hubbard parameters}

The lattice depths were calibrated by modulating the individual lattice beam intensities and observing parametric excitations from the ground to the second-excited Bloch band. We determine the central lattice depth in the center of the trap from the observed parametric resonance frequency with an accuracy of $3\%$.

Our calculation of the on-site interaction energy $U$ is based on the analytical solution for the energy of two atoms in an axially symmetric harmonic potential interacting through an $s$-wave $\delta$-pseudopotential \cite{Idziaszek2006}. Since the actual sinusoidal potential around a single lattice site approaches a harmonic potential only for very deep lattice depths, we employ a (lattice-depth dependent) anharmonic correction factor \cite{Schneider2009} that rescales the analytical solution for the harmonic potential to match the perturbative solution for the sinusoidal potential around vanishing scattering length. This approach has been confirmed experimentally for the bosonic Hubbard model \cite{Mark2011}. Further, we find very good agreement of calculated interaction shifts between different hyperfine-state pairs and the values observed in RF spectroscopy.

In determining the uncertainty of $U/t$ we take into account the uncertainties in the three lattice depths, in the homogeneous magnetic field $B$ and in the scattering length $a_s(B)$. The dominant contribution is the uncertainty in the parametrisation of the Feshbach resonance at around 202~G.

Due to the Gaussian envelope of the lattice beams, the depths of all three lattices $V_{x,y,z}$ vary in the $xy$-plane as one moves away from the trap center. Thus, the in-plane tunnelling $t$ and the on-site interaction $U$ are spatially dependent. Moreover, the tunnelling becomes anisotropic and $t_{x,y}$ have to be considered separately away from the trap center. While we do not take this variation into account in our analysis, we give a quantitative estimate of this variation in the following.

First, we consider the variation along the lines where $V_x = V_y$. There, we have $t_x = t_y$ and the potential at a lattice site is axially symmetric. With the beam waists given in the previous section these lines correspond approximately to diagonals in the $xy$-plane. The radial variation of $U/t$ along these lines can be approximated by 
$(U/t)(r) = (U/t)_0 \left[ 1 - 2.54 \times 10^{-5}\, (r/a)^2 + 2.88 \times 10^{-10}\, (r/a)^4 \right]$ with the lattice constant $a$. This corresponds to a $(5,10,15,20)$~\% reduction compared to $(U/t)_0$ at a radial distance of $r/a=(45,64,80,93)$, respectively. The increase in tunnelling can be approximated by $t(r)/t_0 = 1 + 1.49 \times 10^{-5}\, (r/a)^2 + 9.51 \times 10^{-9}\, (r/a)^3 $, which gives an increase of $(5, 10, 15, 20)$~\% at $r/a = (57,80,97,112)$, respectively. In general, the potential at a lattice site away from the trap center is no longer axially symmetric, and our method to calculate $U$ outlined above is no longer applicable.

\subsection*{Data analysis}
The data was taken by repeatedly scanning the RF frequency of the tomography pulse across the central layer which was placed into the focus of the imaging system. Post-selection of those images where the tomography pulse was resonant with the central layer has been performed according to the following procedure: We extract from every individual image the isopotential averages $n_{S/D}(-V)$ (binned in steps of $\Delta V = h\times \unit[100]{Hz}$) for singles ($S$) and doubles ($D$) and fit the data with NLCE \cite{Khatami2011} (or non-interacting) theory $\eta \times n^\mathrm{theo}_\mathrm{S/D}(\mu_0-V,T)$ with fit parameters $\mu_0$, $T$ and $\eta$, the latter accounting for the efficiency of the tomography pulse. The histogram of the obtained efficiency values $\eta$ exhibits a clear peak at 1. We confirmed that our data for temperatures above $k_B T\sim 2t$ can be equally well described by second-order high temperature series expansion (HTSE) of the Hubbard model. However, the HTSE fits are not able to describe our low-temperature data due to the occurrence of oscillations around filling $n=0.25$ and $n=0.75$ \cite{DeLeo2011}.

In the further data analysis only those images with $0.95 \leq \eta \leq 1.05$ are taken into account. Averaging over multiple experimental repetitions is then performed on the individual density distributions $n(\mu)$. This approach eliminates shot-to-shot fluctuations of the total atom number (which are below $10\%$) from the averaged equation of state data. In Figure 1 of the main text, however, we averaged the data, obtained by the same selection method, in real space and for the shown density profiles applied a Gaussian-shaped spatial filter matched to our imaging resolution.

The compressibility data shown in Figure 2b of the main text is deduced from the derivative of local (second-order) polynomial fits to the equation of state data in a chemical potential range of $h\times \unit[700]{Hz}$. The error bars indicate the standard error of the fit including the error of our determination of $\mu_0$. 

For the extraction of the compressibility data at half-filling shown in Figure 3b of the main text, we fit the equation of state data near half-filling with a third-order polynomial of odd symmetry over a chemical potential range of $h\times \unit[700]{Hz}$ for the low-temperature data and $h\times \unit[1300]{Hz}$ for the high-temperature data. We then choose in a range of $\pm h \times\unit[100]{Hz}$ around the half-filling point $\mu=U/2$ the polynomial fit with the smallest slope and extract from this the compressibility at half-filling.

\begin{acknowledgments}
We thank J. L. Bohn, C. F. Chan, E. Gull, E. Khatami, C. Kollath, J. Leblanc, and M. Rigol for discussion. The work has been supported by BCGS, DFG, the Alexander-von-Humboldt Stiftung, EPSRC and ERC.
\end{acknowledgments}

\end{document}